\documentclass{evn2004}
\input{psfig}
\usepackage{txfonts}
\def\HI{H{\,\small I}}
\newcommand{\kms}{$\,$km$\,$s$^{-1}$}
\newcommand{\mJybeam}{mJy beam$^{-1}$}

\newcommand{\msun}{{$M_\odot$}}
\newcommand{\ltsima} {$\; \buildrel < \over \sim \;$}
\newcommand{\gtsima} {$\; \buildrel > \over \sim \;$}
\newcommand{\lta} {\lower.5ex\hbox{\ltsima}}
\newcommand{\gta} {\lower.5ex\hbox{\gtsima}}
\newcommand{\matHI}{\rm H{\hskip 0.02cm\scriptscriptstyle I}}
\newcommand{\tspin}{$T_{\rm spin}$}
\newcommand{\atms}{atoms cm$^{-2}$}

\begin{document}
   \title{Probing the nature of the ISM in Active Galactic Nuclei through
H{\sc i} absorption
           }
   \author{Raffaella Morganti 
          }

   \institute{Netherlands Foundation for Research in Astronomy, Postbus 2,
              NL-7990 AA, Dwingeloo, NL; morganti@astron.nl
             }

   \abstract{ The physical and kinematical conditions of the gas
   surrounding an active galactic nucleus (AGN) offer key diagnostics
   for understanding the processes occurring in the inner few kpc
   around the nucleus.  Neutral hydrogen can give important insights
   on these regions.  Apart from probing the presence of gas in
   relatively settled conditions (i.e. circumnuclear disks/tori) it
   can also trace the presence of extreme outflows.  Some examples of
   these phenomena are briefly presented.  For the study of the
   neutral hydrogen around AGN the high resolution offered by the VLBI
   is crucial in order to locate the regions where the absorption
   occurs and to study in detail the kinematics of the gas. Recent
   VLBI results are discussed here. }

   \maketitle
\section{Introduction}

The presence of neutral hydrogen in the region surrounding the active
galactic nuclei (AGN) is known since many years. This gas can be
studied via absorption detected against the strong continuum source
(see e.g.  Heckman et al. 1983; van Gorkom et al. 1989; Morganti et
al.  2001, 2002; Vermeulen et al. 2003) and it is now known to be
associated with different structures.  

Neutral hydrogen can be found in tori very close to the AGN.  Although it has
been generally assumed that the tori are composed of dusty molecular clouds,
it is now clear that, under certain conditions, they can be partly formed by
atomic hydrogen (Maloney, Hollenbach\& Tielens 1996).  The \HI\ can also be
associated with larger scale circumnuclear disks (with size ranging from 0.1
and 1 kpc).  These structures are similar to the nuclear optical disks
detected in a large number of early-type galaxies (both radio-loud and
radio-quiet).  These disks (mainly detected by HST) can be seen either in
ionized gas or through their strong dust absorption (van der Marel 2001,
Capetti et al.  2000 and ref.  therein). 

However, the neutral hydrogen can also be associated with more
disturbed structures, like bridges or tails left over from recent
mergers.  The origin of activity in galaxies is often explained as
triggered by merger and/or interaction processes. \HI\ is often seen
associated to all these phenomena.  The idea of merger is supported by
morphological and kinematical evidence (e.g.  Smith \& Heckman 1989,
Tadhunter et al.  1989, Baum et al.  1992).  Torques and shocks during
the merger can remove angular momentum from the gas in the merging
galaxies and this provides injection of substantial amounts of
gas/dust into the central nuclear regions (see e.g. Mihos \& Hernquist
1996).  It is, therefore, likely that in the initial phase of an AGN,
this gas, including atomic hydrogen, still surrounds -- and possibly
obscurs -- the central regions.  AGN-driven outflows have powerful
effects on this dense ISM.  Surprisingly, the neutral hydrogen has
been recently found associated also with such fast outflows (up to
2000 \kms). This finding gives further information on the physical
conditions of the gas in the environment of AGN.  Gas outflows
generated by the nuclear activity are particularly important because
of the effects they can have on the interstellar medium (ISM). This
feedback can be extremely important for the evolution of the galaxy,
up to the point that it could limit the growth of the nuclear
black-hole (e.g.\ Silk \& Rees 1998, Wyithe \& Loeb 2003).

All the above illustrates how important is the gas in the study of
AGN.  The study of the neutral hydrogen is complementary to the
studies of the other phases of the gas - molecular and ionized - in
these regions.  Here, I briefly discuss some of the most recent results
in this area.  While the detection of the \HI\ is usually done with
arcsec resolution observations, the VLBI follow up is crucial in order to be
able to understand in wich of the above mentioned structure the
gas is located and to derive its physical parameters.

\begin{figure}
\centerline{
\psfig{figure=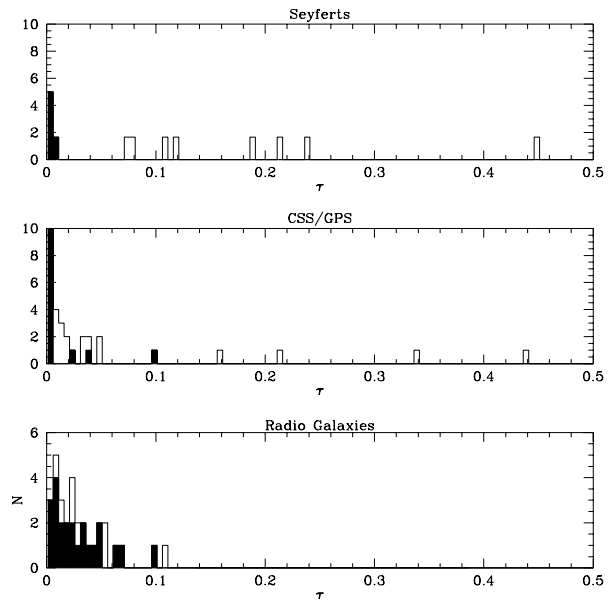,angle=0,width=8cm}}
\caption{ Histograms of the distribution of optical depth ($\tau$) for Seyfert 
galaxies (Gallimore et al. 1999), radio galaxies (Morganti et
al. 2001, Morganti et al, in prep) and CSS/GPS radio sources
(Vermeulen et al. 2003, Morganti et al. 2001). Filled regions indicate
upper limits. }
\end{figure}

\section{Detection of \HI\ in absorption in AGN}

Radio galaxies and radio loud Seyferts have been the subject of many
\HI\ studies.  As result, we know now that about 10-20\% of radio
galaxies show \HI\ absorption against their nuclei while this fraction
goes up to more than 60\% for Seyfert galaxies (Gallimore et
al. 1999).  A group of objects where the fraction of detected
absorption is particularly high are Compact Steep Spectrum (CSS) and
Gigahertz Peaked-Spectrum (GPS, O'Dea 1997) sources. Vermeulen et
al. (2003) found about 50\% of detections in these objects.
The depth of an absorption line ($\Delta S$) depends on the optical
depth ($\tau$), the continuum flux density ($S$) and the covering
factor $c_{\rm f}$ as $\Delta S = c_{\rm f} S (1-e^\tau)$. Usually, a
covering factor $c_{\rm f} = 1$ is assumed.  The distribution of
optical depth ($\tau$) for these three groups of AGN is shown in Fig.~1.

The sensitivity of present days radio telescopes is the main limitation for
the study of the \HI\ absorption.  The typical optical depth observed in the
detected objects is $\tau \sim 0.01 - 0.05$ (i.e.  the flux absorbed by the
\HI\ is few \% of the radio continuum of the source at the frequency of the
redshifted \HI).  Thus, for a typical observation with a noise level in every
channel of about 0.5 \mJybeam\ (1-$\sigma$), these values of optical depth
(detected at 3-$\sigma$ level) can be obtained if the continuum is of the
order of $\sim 50 - 100$ mJy.  It is clear that this is a major limitation for
the study of the \HI\ (e.g.  in weak radio sources or sources with weak radio
cores).  Objects with optical depth $\tau \sim 0.10-0.20$ or larger do exist
but they are rare. 

The histograms of Fig.~1 show that the detected Seyfert galaxies have,
on average, higher optical depth than radio galaxies. CSS/GPS appear
more often detected: the high flux typical of these objects allow to
reach very low optical depth.  On the other hand, the typical core
flux of a radio galaxy is seldom strong enough to reach these limits.
Therefore, the higher detection rate of CSS/GPS is affected by this
bias.  However, it is also the case that objects with high optical
depth are missing from the radio galaxies while are observed among
CSS/GPS.

In particularly bright radio sources, we are, however, in the position
to look for \HI\ with very low optical depth ($\tau \sim 0.001$).
Recent observations making use of the broad band (20~MHz) now
available, e.g. at the upgraded WSRT, have shown that the kinematics
of this gas can be very extreme. Very broad \HI\ absorption features
have been discovered in this way. This will be discussed in Sec.~5.3.

The column densities of the neutral hydrogen follow from $N_{\matHI} =
1.83 \cdot 10^{18} T_{\rm spin} \int \tau dv$ where \tspin\ is the
spin temperature in Kelvin and $v$ is the velocity in \kms. Assuming
the canonical $T_{\rm spin} = 100 $ K, the column densities typically
found are in the range from few times 10$^{19}$ \atms\ to few times
10$^{21}$ \atms.  It should be noted, however, that these values of
$N_{\matHI}$ are likely to be lower limits. In fact, in some physical
situations, such as close to the nuclei of active galaxies or in
outflows, the spin temperature is likely to be as large as a few 1000
K (Maloney et al.1996).

\section{Accurate systemic velocities}

For a proper interpretation of the kinematics of the \HI\ absorption,
it is important to have the systemic velocity of the galaxy as
accurate as possible . Although this sounds trivial, it is not always
an easy information to have available (at least at the level of
accuracy required for the comparison with \HI\ data). It has been often pointed out
(Mirabel 1989, Morganti et al. 2001) that the systemic velocity derived
from emission lines can be both uncertain and biased by motions of the
emitting gas.  This is indicated, in the most extreme situation, by the
detection  in few objects of two different redshift
systems, one derived from the low and the other from the high ionization
lines. Probably the best example of this has been observed in the
radio galaxy PKS~1549-79 (Tadhunter et al. 2001). In this object, the
[OIII]5007\AA\ lines appeared to be associated to gas kinematically
disturbed, likely an outflow due to the interaction with the radio
plasma.

In the southern radio galaxy PKS~1814-63 (see Fig.~2), only the
detailed analysis of the optical spectrum has shown that the emission
lines are actually made of two components: a narrow one that appears
to trace the more extended and quiescent and,therefore, more likely to
define the systemic velocity and a broader component, likely
originating from gas interacting with the radio plasma.

Similar situations have been found in other radio galaxies
(e.g. 4C12.50, Holt et al. 2003; 3C~293, Emonts et al. in
prep.). Interestingly, these are often radio sources where outflows
also associated with the neutral hydrogen are detected (see Sec.~5.3).
This further emphasize the importance of study both the ionized and
the neutral component of the gas in order build a more complete (and
correct) picture of the physical conditions around AGN.

\begin{figure}
\centerline{
\psfig{figure=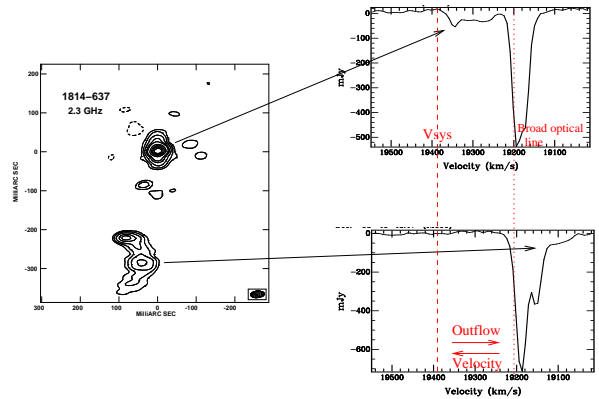,angle=0,width=8cm}}
\caption{LBA observations of the southern radio galaxy PKS~1814-63 
(Tzioumis et al. in prep.). The nucleus is tentatively identified with
the northern component. A blueshifted shallow component of the
absorption is detected. The most blueshifted component appear to be
located against the southern component. In order to identify this as
an outflow, new accurate measurments of the systemic velocity was
needed (from optical data, Holt et al. in prep). The new value,
derived from the narrow component in the optical emission lines, is
indicated with the dashed line. The dotted line indicate the velocity
of the broad component detected in the optical emission lines. }
\end{figure}

\section{Circumnuclear tori and disks}

As described above, \HI\ gas can be associated with
circumnuclear disks and tori. To establish whether the \HI\ absorption
is coming from these structures is not always easy. Often, given the
limited size of the underlying continuum, clear kinematical signatures
of a rotation cannot be seen. Thus, one of the criteria to distinguish between
these structures is the width of the absorption line. While the
\HI\ associated with circumnuclear disks show relatively broad
absorption (typically $> 150$ \kms), \HI\ associated with larger scale
structures is usually observed as narrow absorption features (see
e.g. the case of Centaurus~A, van Gorkom et al. 1986).

In Seyfert galaxies, Gallimore et al. (1999) found that, with the
exception of NGC~4151, the absorbing gas traces 100 pc-scale rotating
disks aligned with the outer galaxy disk.  In NGC 4151, \HI\
absorption measurements using MERLIN and VLBA indicate a torus $\sim$ 70 pc
in radius and $\sim$ 50 pc in height (Mundell et al. 1995, 2003).  A cartoon
illustrating the geometry of the system is shown in Fig.~3.

\begin{figure}
\centerline{
\psfig{figure=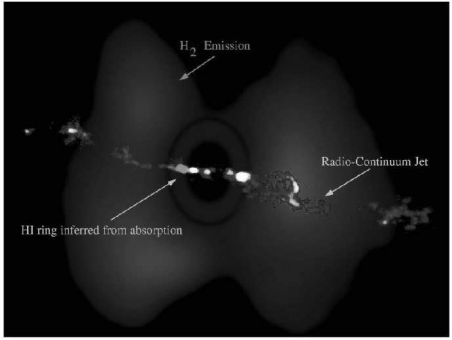,angle=-90,width=8cm}}
\caption{Montage of the inner 250 pc of NGC 4151 from Mundell et al. (2003), 
showing torus of H$_2$ emission in green (from Fernandez et al. 1999), ring of
\HI\ inferred from absorption measurements in blue and 1.4 GHz radio continuum
emission from radio jet in red. Ionized gas (black) is assumed to fill the 
torus inside the \HI\ ring. The segments missing to the north and south of the 
H$_2$ torus are due to the limited filter width that excluded the high-velocity 
wings of the H$_2$ line; this provides evidence for rotation of the torus as 
the northern and southern segments contain the gas with the highest radial 
velocities (i.e., line wings) if the torus is rotating. }
\end{figure}

\begin{figure*} \centerline{
\psfig{figure=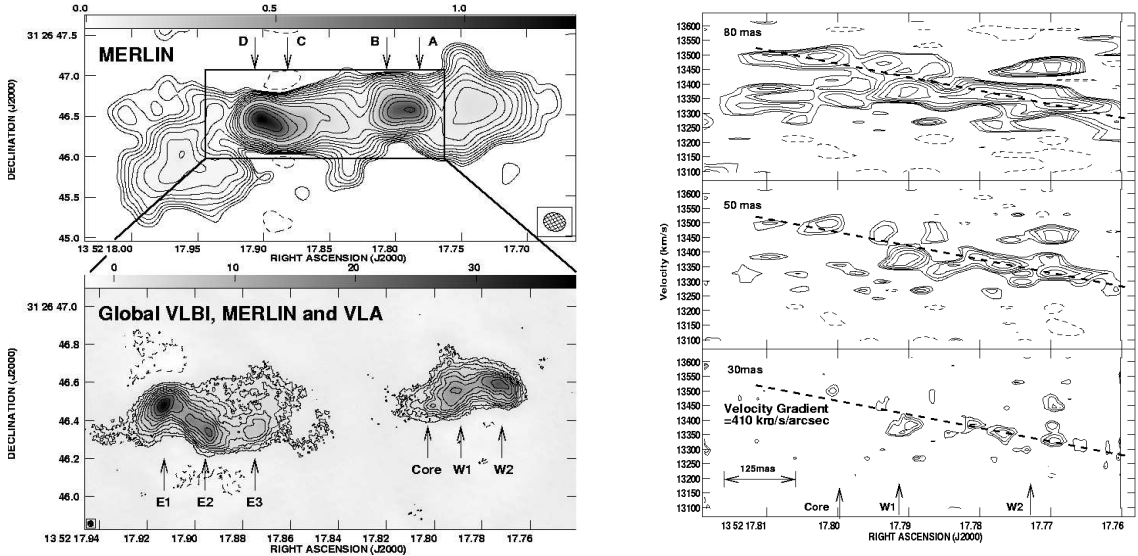,angle=0,width=15cm}} \caption{{\sl (Left)}
Sub-arcsecond continuum structure of the inner few kiloparsecs of 3C\,293
(from Beswick et al.  2004).  The top contour map shows the 1.359 GHz radio
continuum structure observed with MERLIN at a resolution of
0\farcs23$\times$0\farcs20.  The lower panel shows the global VLBI, MERLIN and
VLA+PT contoured image of the inner jet of 3C\,293 with angular resolution of
30\,mas.  This map is contoured at multiples of $\sqrt2$ times
1.3\,mJy\,beam$^{-1}$.  {\sl (Right)} Multi-resolution position-velocity plots
of H{\sc i} absorption against the western jet component at the centre of
3C\,293 (from Beswick et al.  2004).  In each of these diagrams the absorption
signal has been averaged over the declination range of the continuum source. 
The dashed line shown on all three plots represents a velocity gradient of
410\,km\,s$^{-1}$\,arcsec$^{-1}$.  The spatial position of radio continuum
components labelled in the figure on the left are also shown by arrows
positioned along the bottom plot.  Images and plots are taken from Beswick et
al.  (2004).} \end{figure*}

Due to the weakness of the radio core in powerful radio galaxies
(i.e. Fanaroff-Riley II), evidence for \HI\ associated with tori has
been found only in few cases. A possible candidate is Cygnus~A. In this
object, a 50 pc-scale, rotating, flattened structure has been found
from the VLBI observations (Conway 1999). However, recent optical
observations have shown that the situation may be more complicated
than this and that the \HI\ absorption includes also the signature of
an inflowing cloud (see Sec.~5.1).

A more clear case comes from the study of the kinematics and
distribution within the central kiloparsec of the \HI\ in the radio
galaxy 3C~293 (Beswick et al. 2004).  Strong H{\sc i} absorption is
detected against the majority of the inner kiloparsec of 3C\,293. This
absorption is separated into two dynamically different and spatially
resolved systems. This result is illustrated in Fig.~4.  Against the
eastern part of the inner radio jet narrow H{\sc i} absorption is
detected and shown to have higher optical depths in areas co-spatial
with a central dust lane. Additionally, this narrow line is shown to
follow a velocity gradient of $\sim$50\,km\,s$^{-1}$\,arcsec$^{-1}$,
consistent with the velocity gradient observed in optical spectroscopy
of ionised gas. The narrow H{\sc i} absorption, dust and ionised gas
appear to be physically associated and situated several kiloparsecs
from the centre of the host galaxy. Against the western jet emission
and core component, broad and complex H{\sc i} absorption is detected.
A possible explanatin for this is that the H{\sc i} is situated in
rotation about the core of this radio galaxy with some velocity
dispersion resulting from in-fall and outflow of gas from the core
region. If this explanation is correct, then the mass enclosed by the
rotating disk would be at least 1.7$\times$10$^9$ solar masses within
a radius of 400\,pc.

Powerful compact (steep spectrum) radio sources are uniquely suited for
investigations into the physics of the central engines, in particular to study
the kinematics of the gas within 100 pc of the core (see e.g.  Vermeulen et
al.  2003).  Pihlstr\"om et al.  (2003) have studied the distribution of the
\HI\ absorbing gas in a sample of these sources.  They find that smaller
sources ($<0.5$ kpc) have larger \HI\ column density than the larger sources
$> 0.5$ kpc) (see Fig.5).  This result can be explained both as a spherical
and an axi-symmetric gas distribution, with a radial power law density
profile, although these authors argue that the disk distribution is the most
likely. 

Evidence of \HI\ associated with cicumnuclear tori has been reported
for some of these compact sources (see e.g. Conway 1997, Peck \&
Taylor 2001).  One of the best examples of this type is the
Compact Symmetric Object (CSO) 1946+708.  The \HI\ absorption in
1946+708 consists of a very broad line and a lower velocity narrow
line which are visible toward the entire $\sim$100 pc of the continuum
source, Peck et al. 1999). with thickness of about 100 pc and column
density of the order of $10^{23}$ cm$^{-2}$ (with T$_{spin}$ of
several thousand K). The broad line has low optical depth and peaks in
column density near the core of the source.  This is consistent with a
thick torus scenario in which gas closer to the central engine is much
hotter, both in terms of kinetic temperature and spin temperature, so
a longer path-length through the torus toward the core would not
necessarily result in a higher optical depth.  The high velocity
dispersion toward the core of 1946+708 is indicative of fast moving
circumnuclear gas, perhaps in a rotating toroidal structure.  Further
evidence for this region of high kinetic energy and column density is
found in the spectral index distribution which indicates a region of
free-free absorption along the line of sight toward the core and inner
receding jet.  The \HI\ optical depth increases gradually toward the
receding jet.  The information derived from the \HI\ can be
particularly useful to constrain characteristics the central torus
when combined with hard X-ray data. This has been done in the case of
two possible Compton-thick galaxies studied by Risaliti et al. (2003).

\begin{figure}
\centerline{
\psfig{figure=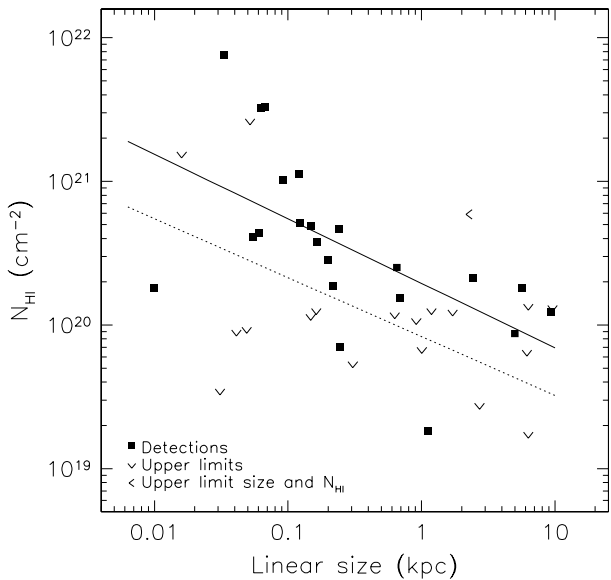,angle=0,width=8cm}}
\caption{Absorbed \HI\ column density versus
projected linear size for CSS and GPS sources. There is an anti-correlation between the source
size and the amount of absorbing gas (from Pihlstr\"om et
al. 2003).}
\end{figure}

\begin{figure}
\centerline{
\psfig{figure=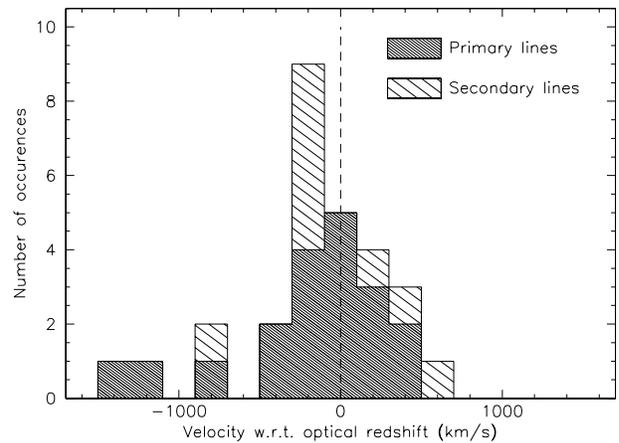,angle=-90,width=8cm}}
\caption{\HI\ line velocities  compared to the optical
velocities for CSS and GPS sources from Vermeulen et al. (2003).}
\end{figure}

In low luminosity radio galaxies the situation could be different.
The high detection rate of optical cores, the lack of large absorption
in X-ray (Chiaberge et al. 1999) and possibly also the relatively low
detection rate of
\HI\ absorption (Morganti et al.\ 2001) suggest that the standard
pc-scale geometrically thick torus is not present in these radio
galaxies.  The presence of thin disks has been claimed from \HI\
observations in the case e.g.\ NGC~4261 (van Langevelde et al.\ 2000).
For this object, the VLBI data suggest that the \HI\ absorption is due
to a disk of only $\sim$1.3 pc thick projected against the
counter-jet.  In NGC~4261, evidence for the presence of such nuclear
disk are also found in HST images. The idea of thin disks can be
investigated in more detail by correlating the presence (or
absence) of \HI\ absorption with the optical characteristics.  This
has been done for a sample of radio galaxies (selected from Capetti et
al.\ 2000) for which information about the presence of optical cores
and nuclear dusty disks/lanes (from HST images) is available.
Interestingly, \HI\ absorption was detected in the two galaxies that
have dust disks/lanes and {\sl no} optical cores.  In these cases, the
column density of the absorption is quite high ($ > 10^{21}$ cm$^{-2}$
for T$_{\rm spin} = 100$ K) and the derived optical extinction $A_B$
(between 1 and 2 magnitudes) is such that it can, indeed, produce the
obscuration of the optical cores.  In the other two cases,
\HI\ absorption has been detected despite the presence of optical cores.
However, the column density derived from the detected absorption is
much lower ($\sim 10^{20}$ cm$^{-2}$ for T$_{\rm spin} = 100 $ K) and
the derived extinction is of the order of only a fraction of a
magnitude. This is, therefore, consistent with what expected if the
circumnuclear disk are thin in these radio galaxies.

\section{Unsettled gas}

\subsection{Any evidence for infall?}

Evidence for infalling gas was reported by van Gorkom et al.  (1989).  In a
sample of radio galaxies, \HI\ absorption was detected either close to the
systemic velocity or systematically redshifted, indicating therefore a
prevalence of gas falling into the nucleus.  This result does not appear to be
confirmed by more recent observations.  For example, the study of \HI\
absorption in compact radio sources by Vermeulen et al.  (2003) shows that
there is evidence for significant gas motions and not only positive but even
more negative \HI\ velocities (up to more than $v = -1000$ \kms\ compared to
the systemic velocity) are found (see Fig.~6).  This is indicating that gas
flowing out of the galaxy is also present.  Indeed, clear cases of fast gas
outflows have now been detected as described below (see Sec.  5.3). 

One of the most promising case of infalling gas was found in the
radio galaxy NGC~315. A very narrow and highly redshifted ($\sim 500$
\kms) \HI\ absorption  was reported by Heckman et
al. (1983) and Dressel et al. (1983). VLBI observations (Peck 1999,
Morganti et al. in prep.) are now showing that this absorption appears
to cover a region of about 9 pc of the source, from the core to the
first part of the jet.  A likely explanation for this absorption is
that of a cloud at large distance from the nucleus (like tidal debris,
Wakker et al.\ 2002) detected, in projection, against the nucleus.
This seems to be more favorable over the possibility of
a small cloud falling into the nucleus and feeding the AGN.

An interesting case of cloud falling into the nucleus has been
recently suggested for Cygnus~A (Bellamy et al. 2004).  Near-IR data
show the existence of an off-nucleus molecular cloud, that is
redshifted respect to the systemic velocity (measured accurately from
stellar features). This suggests the presence of a giant molecular
cloud falling though the ``heart'' of Cygnus~A. Interesting, the
redshift of this cloud is in agreement with that of the \HI\
absorption (or part of it) indicating, therefore, that the two
phenomena may be linked.

\begin{figure*}
\centerline{
\psfig{figure=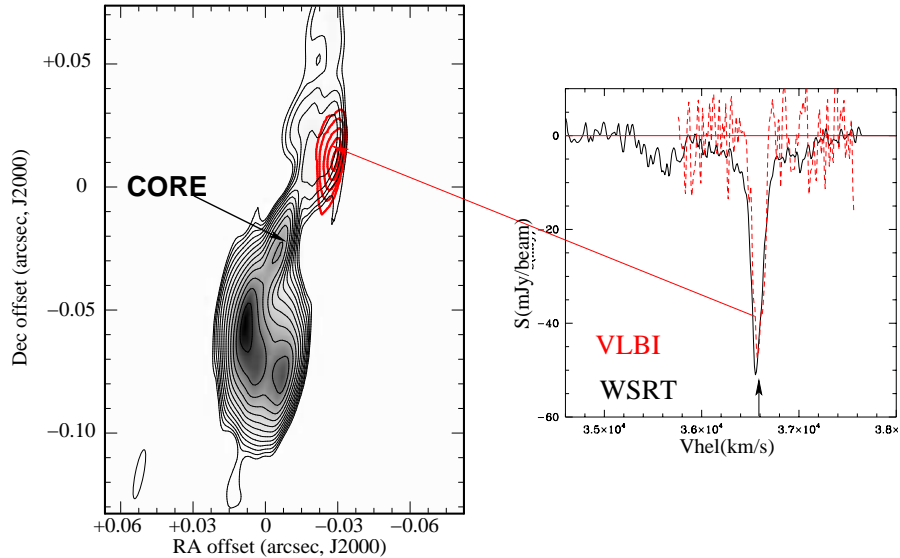,angle=0,width=13cm}}
\caption{{\sl (Left)} VLBI continuum image (grey scale and thin contours) of 
4C~12.50 (from Morganti et al. 2004a) superimposed onto the total
intensity of the (narrow) \HI\ absorption observed at the systemic
velocity (thick contours). The position of the radio core is also
indicated. The contour levels for the continuum image are: 5 \mJybeam\
to 800 \mJybeam\ in steps of factor 1.5. {\sl (Right)} \HI\ absorption
profile observed with the WSRT (black) and VLBI (red). A broad,
shallow \HI\ absorption is detected in the WSR Tobservations. Due to
the narrower band, this broad absorption is not detected in the VLBI
observations. }
\end{figure*}

\subsection{Gas cocoon around AGN}

As mention in the introduction, \HI\ absorption can also trace gas
distributed in a more complex way around the AGN.
In the low luminosity active galactic nucleus, NGC~1052, the VLBI
study of the \HI\ has revealed atomic gas in front of the approacing
jet as well as the receding jet (Vermeulen et al. 2003b). The gas
appeared to be associated with three velocity systems. One system can
be as close as 1-2 pc from the core. The other systems could be local
to the AGN environment or distributed on galactic scales.

\HI\ absorption can be used to trace a particularly rich medium 
that is characteristics of some radio galaxies -- perhaps those resulting
from major mergers or in which a merger happened not so long ago.  One
example is the radio galaxy 4C~12.50 (see Fig.~7, Morganti et
al. 2004a), a galaxy that has often been suggested to be a prime
candidate for the link between ultraluminous infrared galaxies and
young radio galaxies. In this object, deep and relatively narrow
\HI\ absorption (observed at the systemic velocity) is associated with an
off-nuclear cloud ($\sim 50$ to 100 pc from the radio core) with a
column density of $\sim 10^{22}\ T_{\rm spin}/(100\ {\rm K}$)
cm$^{-2}$ and an \HI\ mass of a few times $10^5$ to $10^6$
\msun. There are more examples of objects where the \HI\ traces the
rich medium surrounding the active nucleus. Examples of off-nuclear
\HI\ absorption are found in 3C~236 (Conway \& Schilizzi 2000) and,
more recently, in the CSO 4C~37.11 (Maness et al.\ 2004) where a broad
($\sim 500$ \kms) absorption line was found in the region of the
southern hot-spot.

This may have important implications for the evolution of the radio
jets. Although this gas will not be able to confine the radio source,
it may be able to momentarily destroy the path of the jet as shown
also by numerical simulations (Bicknell et al.\ 2003). Thus, this
interaction can influence the growth of the radio source until the
radio plasma clears its way out.

\subsection{Fast Outflows}

Gas outflows associated with active galactic nuclei (AGN) provide
energy feedback into the ISM that can profoundly affect the evolution
of the central engine as well as that of the host galaxy. The mass-loss
rate from these outflows can be a substantial fraction of the accretion
rate needed to power the AGN.

Fast outflows have now been detected in a large fraction of nearby AGN
via observations at visible, X-ray and UV wavelengths associated to ionized gas (see
e.g. Veilleux et al. 2000, Kriss 2004, Elvis, Marengo \& Karovska
2002 and ref. therein for an overview).  It
is, therefore, not too surprising to find such outflows also in radio
galaxies (see Morganti et al.  2003a for a summary of recent results).
However, it is extremely intriguing the discovery of a number of radio
sources where the presence of fast outflows (up to 2000 \kms) is associated not only
with ionized but also with {\sl neutral} gas.  This finding gives new
and important insights on the physical conditions of the gaseous
medium around an AGN.  The best examples so far are the radio galaxies
3C~293 (Morganti et al.  2003b, see Fig.8a) and 4C~12.50 and the Seyfert galaxy
IC~5063 (Oosterloo et al.  2000).  It is also worth noticing that
outflows of ionized gas are also associated with these neutral
outflows (see Morganti et al.  2003a).

An other interesting object where an \HI\ outflow has been detected is
the Compton-thick, broad-line and GPS radio galaxy OQ~208. The \HI\
spectrum of this source (that is only 10 pc in size) is shown in
Fig.~8b.  Guainazzi et al. (2004) suggest that in this source we
coulld be seeing the jets piercing their way through a Compton-thick
medium pervading the nuclear environment.  The outflow detected in
\HI\ (see Fig.8b) would be an other indication of this process.
Guainazzi et al. (2004) also suggest that if the jets have to interact
with such a dense medium, one could largely underestimating the radio
activity dynamical age determinated for this kind of sources from the
observed hot-spot recession velocity. A similar situation could be
occuring the the GPS 4C12.50 described above.

A number of possible hypotheses can be made about the origin of the
gas outflow (e.g., starburst winds, radiation pressure from the AGN,
adiabatically expanded broad emission line clouds). In some cases, 
the possibility that they are 
 driven by the interaction of the
radio jet with the ISM seems to be favored.  To investigate whether this is
indeed correct, high-resolution (VLBI) studies are in progress to find
the exact location of the ouflowing gas. So far these outflows
have been found in objects that are either in the early-stage of their
evolution (like 4C~12.50) or, perhaps, in a phase of re-started
activity (as might be the case for 3C~293).  Another characteristic of
these galaxies is the presence of a ``young'' stellar population (from
their optical spectra, see e.g. Tadhunter et al.\ 2002).  Such a
component (with ages between 0.5 and 2 Gyr) can be considered an
indication that the galaxy is indeed in a  stage of its
evolution, when large amounts of gas/dust - likely from the merger
that triggered the activity - are still present in the inner region
and the radio jet is strongly interacting with it.   A more
systematic search for fast gas outflows in radio galaxies is now in
progress  and has already revealed more cases of broad,
blueshifted \HI\ absorptions.

\begin{figure*}
\centerline{
\psfig{figure=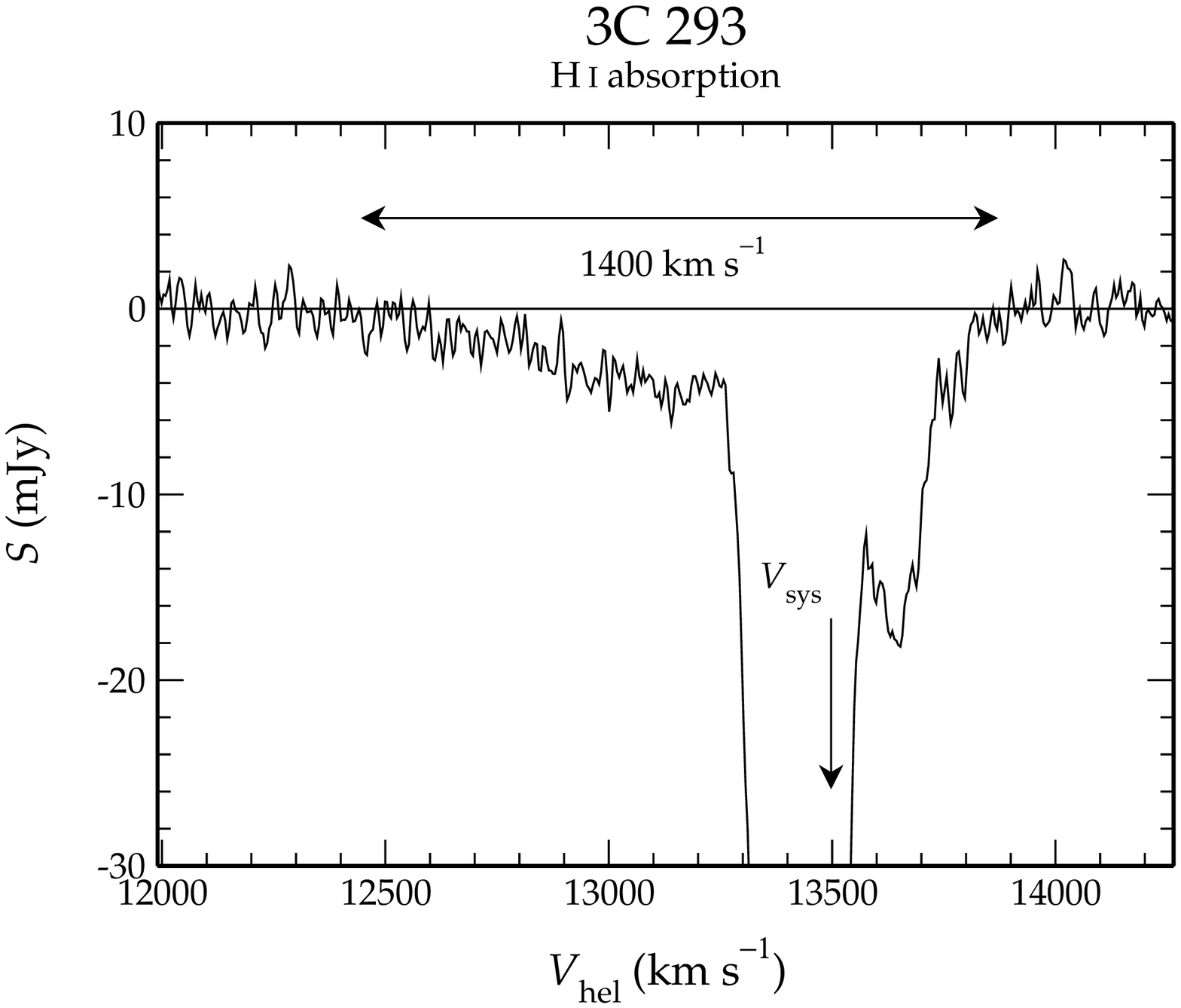,angle=0,width=7cm}
\psfig{figure=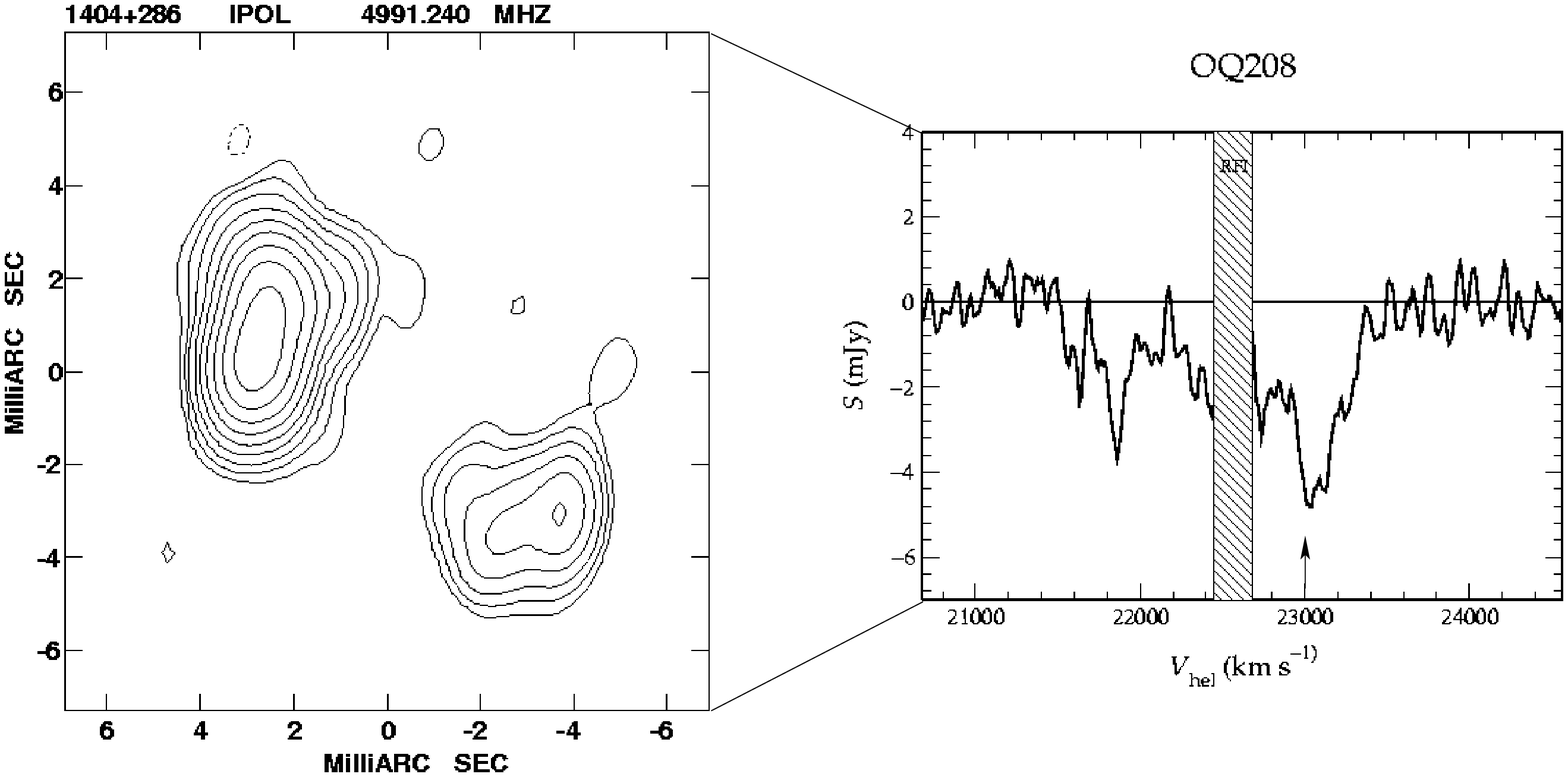,angle=0,width=9cm}}
\caption{{\sl (Left)} The \HI\
absorption profile detected in 3C~293 from the WSRT observations. The
spectrum is plotted in flux (mJy) against optical heliocentric
velocity in \kms\ (from Morganti et al. 2003). {\sl (Right)} Broad \HI\
absorption detected (using the WSRT) against the compact source OQ~208. The systemic
velocity derived by Marziani et al. (1997) is also indicated. The VLBI
radio continuum (that is only $\sim 10$ pc in size) is taken from
Stanghellini et al. (1997).}
\end{figure*}

\section{Conclusions}

The \HI\ is an important tool to study the physical conditions of
the gas around AGN. Although the main limitation of these studies is
the sensitivity of present day radio telescopes, the 
possibility now becoming more and more available of broad band observations
allows to explore the presence of
kinematically disturbed \HI. This may represent a relatively
common phenomena, perhaps related with the evolutionary stage of the
radio sources.  The importance of understanding the physical conditions
of the gas in the environment of the AGN (both in the circumnuclear
tori as well as in the AGN-driven outflow) is illustrated by the
interest and the wealth of observations performed, e.g., in the
optical and X-ray bands.  However, the possibility of imaging this gas
at very high resolution -- by obtaining \HI\ absorption with
milli-arcsec resolution -- is unique to the radio band and the VLBI
technique. This combined with the broad band (i.e. to instanteneously
cover up to $\sim 4000$ \kms) is providing an extremely powerful tool to
investigate the conditions (including the extreme one that now we know
can exist) of the atomic gas around AGN.

The next step is, however, the dramatic improvement and possibilities
that the new generation of radio telescope will offer. A summary of
the possibilities that the Square Kilometer Array will open for the study of
th eenvironment of the AGN are
summarized in Morganti et al. (2004b).  Apart from the more detailed
study of the \HI\ in single objects, the SKA will provide the
possibility of performing large surveys and understand the occurence
of the phenomena described above and their relation with the
properties of the host galaxy. The SKA will allow to investigate an
unexplored
region of parameter space. 
At present, serendipitous \HI\ surveys can already be carried out in
every deep field, for example by using 
spectral-line mode, in which continuum observations (see e.g. Morganti
et al.  2004c for the case of the Spitzer Space Telescope First-Look
Deep survey done with the WSRT).

With the high sensitivity expected from the SKA, we will be able to search for
\HI\ absorption at $\tau \sim 0.01$ level (the typical absorption found in
cases of circumnuclear tori) on every source stronger than only a few mJy of
any observed field.  It will be like searching every source of the NVSS
catalogue for \HI\ absorption.  The large instantaneous bandwidth will ensure
that a large range in redshift is covered to detect this absorption.

\end{document}